\documentclass[preprint,aps,prc,showpacs,nofootinbib,secnumarabic]{revtex4-1}

\usepackage{graphicx}% Include figure files
\usepackage{bm}% bold math
\usepackage{color}

\begin{document}

\title{Charmonium production in antiproton-nucleus reactions at low energies}

\author{A.B. Larionov$^{1,2}$\footnote{Corresponding author.\\ 
        E-mail address: larionov@fias.uni-frankfurt.de},
        M. Bleicher$^{1,3}$, A. Gillitzer$^4$, M. Strikman$^5$}

\affiliation{$^1$Frankfurt Institute for Advanced Studies (FIAS), 
             D-60438 Frankfurt am Main, Germany\\ 
             $^2$National Research Center "Kurchatov Institute", 
             123182 Moscow, Russia\\
             $^3$Institut f\"ur Theoretische Physik, J.W. Goethe-Universit\"at,
             D-60438 Frankfurt am Main, Germany\\  
             $^4$Institut f\"ur Kernphysik, Forschungszentrum J\"ulich, D-52425 J\"ulich, Germany\\
             $^5$Pennsylvania State University, University Park, PA 16802, USA}

\date{\today}

\begin{abstract}
The $J/\Psi(1S)$ and $\Psi^\prime(2S)$ production near threshold
in antiproton-nucleus reactions is calculated on the basis of the Glauber model. 
The model takes into account the antiproton (pre)absorption,
proton Fermi motion, and charmonium formation length.
We confirm earlier prediction that the charmonium production 
in $\bar pA$ collisions at $p_{\rm lab}=3-10$ GeV/c is not influenced by formation length effects 
and is very well suited to determine the genuine charmonium-nucleon dissociation cross sections.
The comparison is performed with the $J/\Psi$ photoproduction at high energies, where 
formation length effects play a very important role.
However, we demonstrate that the detailed structure of the proton and neutron density profiles 
has to be taken into account, if one wants to extract information on the $J/\Psi N$ 
dissociation cross section from $J/\Psi$ transparency ratios.
These studies are relevant for the upcoming PANDA experiment at FAIR.    
\end{abstract}

\pacs{25.43.+t;~14.40.Lb;~24.10.Ht;~25.20.Lj;~24.10.Jv}

\maketitle

\section{Introduction}

The exploration of the properties of quantum chromodynamics (QCD) matter under extreme
conditions is one of the most challenging endeavors of todays high-energy physics.
The dynamics of heavy quarks, i.e. $c$- and $b$-quarks, and their bound states has
been shown to provide new insights into these questions.  
A better understanding of the interactions of charmonia with cold nuclear matter
is especially very important for the studies of charm production in heavy-ion collisions, 
in particular, of the $J/\Psi$ suppression in a quark-gluon plasma \cite{Matsui:1986dk}.
The interpretation of $J/\Psi$ yields from p-A and non-central AA collisions
at moderate energies ($\sqrt{s}=17.3$ GeV) in terms of a usual hadronic 
scenario requires the $J/\Psi$ dissociation cross section on
a nucleon to be about $6-7$ mb \cite{Gerschel:1993uh,Kharzeev:1996yx}. 
Moreover, as stated in Ref. \cite{Gerschel:1993uh}, such cross sections are consistent 
with the world data on $J/\Psi$ transparency ratios from other reactions induced by 
elementary particles ($\gamma,~\pi,~\bar p$) on nuclei.
However, if one takes into account contribution of $J/\Psi$ produced in the decays of 
higher charmonium states and larger cross section of inelastic $\chi_c N$ interactions, 
the $J/\Psi N$ dissociation cross section turns out to be about two times smaller, 
i.e. about 3.5 mb \cite{Gerland:1998bz}. 

A problematic feature of all existing experimental data on
$J/\Psi$ transparency ratios,
\begin{equation}
   S_A = \frac{\sigma_{p A \to J/\Psi X}}{A\sigma_{p N \to J/\Psi X}}~,  \label{S_A}
\end{equation}
(here "$p$" denotes any kind of elementary projectile and $N$ stands for the nucleon) 
in elementary particle-induced reactions is that the charmonia are produced 
at high momenta from about 20 GeV/c up to several TeV/c. 
Due to the large Lorentz-boost of the charmonia,
the existing data only give access to the interactions of prehadronic $c \bar c$
and $c \bar c g$ configurations with the nuclear medium which would transform 
into $J/\Psi$ well after the target.
Thus, the extracted value of the dissociation cross section needs to be studied in other 
kinematical situations, when the charmonium moves more slowly through the nuclear target.

Brodsky and Mueller \cite{Brodsky:1988xz} proposed to measure color transparency effects for
$J/\Psi$ production in $\bar p$-nucleus interactions, where the threshold beam momentum 
($p_{\rm thr}=4.07$ GeV/c for $J/\Psi$ production in $\bar p p$ interactions) is quite low.
The first theoretical study of this reaction has been done
in Ref. \cite{Farrar:1989vr} within the color diffusion and Glauber models taking into account
the Fermi motion of the nucleons. The most important results of Ref. \cite{Farrar:1989vr} are
(i) the high sensitivity of the production cross section $\sigma_{\bar p A \to R (A-1)^*}$
of a charmonium state $R$, where $R$ stands for $J/\Psi,~\Psi'$ or $\chi_c$ on a nucleus 
to the $RN$-dissociation cross section;\\
(ii) due to the Fermi motion the cross section of charmonium  production on a nucleus is 
strongly suppressed, i.e. $\sigma_{\bar p A \to R (A-1)^*} \sim (10^{-4}-10^{-3}) Z \sigma_{\bar p p \to R}$
at the beam momentum of the on-shell $R$ production\footnote{This estimate corresponds to the
perfect beam resolution for $\bar p p$ collisions. See Eq.(\ref{estim}) below and the text after it.};\\
and (iii) $\sigma_{\bar p A \to R (A-1)^*}$ is sensitive to the color transparency effect for the 
incoming antiproton (which can, however, be reformulated as a phenomenological treatment
of $\bar p$ absorption), but insensitive to the color transparency for the produced  
$J/\Psi$ and having modest sensitivity for heavier charmonium state $R$.

Later-on, in work \cite{Gerland:2005ca}, the $J/\Psi$ and $\Lambda_c$ production in $\bar p A$ collisions 
at the $\Psi'$ and $J/\Psi$ production thresholds has been addressed with a focus on the effects of
the non-diagonal transitions $\Psi' N \to J/\Psi N$. However, Fermi motion effects have been neglected
in Ref. \cite{Gerland:2005ca}.
We would also like to mention a very inspiring feasibility study for Fermi National Accelerator Laboratory
(Fermilab) \cite{Seth:2001wq},
which for the first time presented the beam momentum dependence of the charmonium production cross sections 
in $\bar p A$ collisions with heavy gas targets (CH$_4$, N, O, Ne, Ar and Xe).

The main purpose of the present work is to perform a detailed theoretical analysis of the 
$J/\Psi$ production in $\bar p A$ collisions near threshold. This is one of the subjects
of the planned PANDA experiment at FAIR \cite{PANDA}.
Comparisons of the $J/\Psi$ transparency ratios in $\bar p A$ and $\gamma A$ reactions
are performed with a focus on the sensitivity to the $J/\Psi N$ dissociation cross section.  

Section \ref{model} explains our model. The model predictions for the charmonium 
production in $\bar pA$ reactions are given in Sec. \ref{results} with an emphasis 
on the sensitivity to the charmonium-nucleon dissociation cross sections. 
For comparison, we have also calculated the $J/\Psi$ transparency 
ratios in photo-induced reactions and showed that the existing experimental data
from Stanford Linear Accelerator Center (SLAC) for $E_\gamma=20$ GeV \cite{Anderson:1976hi} 
and, moreover, of  Fermilab for $E_\gamma=120$ GeV \cite{Sokoloff:1986bu} do not allow 
to constrain the $J/\Psi N$ 
dissociation cross section due to the formation length effects and large experimental errors. 
Finally, in order to evaluate the influence of multistep processes on
$J/\Psi$ production in $\bar p A$ reactions 
the results obtained within the Giessen Boltzmann-Uehling-Uhlenbeck (GiBUU) model \cite{Buss:2011mx}
are presented.  
The conclusions are given in Sec. \ref{conclusions}.

\section{Model}
\label{model}

In the reaction $\bar p A \to R + X$ at beam momentum close to the charmonium $R$ production 
threshold, the produced charmonium carries nearly the entire antiproton momentum.
Therefore, we can apply a Glauber model similar to that of Ref. \cite{Farrar:1989vr}. 
The cross section
of the charmonium $R$ production in a $\bar p A$ collision is given by
\begin{equation}
   \sigma_{\bar p A \to R (A-1)^*}=2\pi\int\limits_0^\infty\,db\, b\,
   v_{\bar p}^{-1}\int\limits_{-\infty}^\infty\,dz 
   P_{\bar p,{\rm surv}}(z,b) \Gamma_{\bar p \to R}(z,b)
   P_{R,{\rm surv}}(z,b)~,                            \label{sigma_pbarA2R}
\end{equation} 
where the integration is done over the antiproton impact parameter $b$ and
the longitudinal coordinate $z$. $v_{\bar p}=p_{\rm lab}/E_{\bar p}$ is 
the antiproton velocity with respect to the target nucleus.
The in-medium width of the antiproton with respect to the charmonium production is 
\begin{equation}
   \Gamma_{\bar p \to R}(z,b)=\int\,\frac{2d^3p}{(2\pi)^3} v_{\bar p p} 
    \sigma_{\bar p p \to R X}(p,p_{\bar p}) f_p(z,b,\mathbf{p})~,        \label{Gamma_pbar2R}
\end{equation}
where $v_{\bar p p}=q\sqrt{s}/E_{\bar p}E_p$ is the antiproton-proton relative velocity
with $q=\sqrt{s/4-m^2}$ being the center-of-mass (c.m.) momentum 
of the antiproton and the proton at the c.m. energy of $\sqrt{s}$; $m=0.938$ GeV
is the nucleon mass. To simplify the discussion we use here the simplest model 
for the proton momentum distribution, i.e. the Fermi distribution 
$f_p(z,b,\mathbf{p})=\Theta(p_{F,p}-|\mathbf{p}|)$,
where $p_{F,p}=(3\pi^2\rho_p(z,b))^{1/3}$ is the local Fermi momentum of protons,
and $\rho_p$ is the local proton density.
The more realistic spectral function with correlations would slightly reduce
the value of cross section near maximum and add momentum tails removing the 
sharp cutoffs.
Currently the program of the experimental studies of the short-range correlations
is under way at the TJNAF and models of the nuclear spectral functions incorporating
these findings are being developed, for a recent review see \cite{Arrington:2011xs}.
This would allow in the near future to perform more accurate calculations of the rates
of the charm production for the kinematics where $\bar p$ produces $J/\Psi$ 
in the interaction with a fast nucleon.

If the beam momentum is close to that of exclusive $R$ production at the mass pole, 
then all processes except $\bar p p \to R$ can be neglected, and one can 
use in Eq.(\ref{Gamma_pbar2R}) the exclusive resonance production cross section 
$\sigma_{\bar p p \to R}$ instead of the inclusive one, $\sigma_{\bar p p \to R X}$ .
For $\sigma_{\bar p p \to R}$ we apply the relativistic Breit-Wigner formula,
\begin{equation}
   \sigma_{{\bar p} p \to R}=\frac{3\pi^2}{q^2}
           \sqrt{s} \Gamma_{R \to {\bar p} p} {\cal A}_R(s)~,  \label{sig_pbarp2R}
\end{equation}
with the resonance spectral function
\begin{equation}
   {\cal A}_R(s)=\frac{1}{\pi} \frac{\sqrt{s} \Gamma_R}{(s-m_R^2)^2+s\Gamma_R^2}~.
                                                 \label{A_R}
\end{equation}

The remaining two important ingredients of Eq.(\ref{sigma_pbarA2R}) are the survival probability
of the antiproton until it reaches the point $(z,b)$,
\begin{equation}
   P_{\bar p,{\rm surv}}(z,b)
  =\exp\left\{-\int\limits_{-\infty}^z\,dz'\rho(z',b) 
\sigma_{\bar p N}^{\rm inel}(p_{\rm lab})\right\}~, \label{P_pbar_surv}
\end{equation}
and the survival probability of the charmonium $R$ until it is emitted to the vacuum
\begin{equation}
   P_{R,{\rm surv}}(z,b)
  =\exp\left\{-\int\limits_z^\infty\,dz'\rho(z',b) 
   \sigma_{RN}^{\rm eff}(p_R,z'-z)\right\}~. \label{P_R_surv}
\end{equation}
In Eqs.(\ref{P_pbar_surv}),(\ref{P_R_surv}), $\rho=\rho_p+\rho_n$ is the total nucleon density,
$\sigma_{\bar p N}^{\rm inel}$ is the $\bar p N$ inelastic cross section
\begin{equation}
   \sigma_{\bar p N}^{\rm inel}(p_{\rm lab})=\sigma_{\bar p N}^{\rm tot}-\sigma_{\bar p N}^{\rm el}~,
                             \label{sigma_pbarN^inel}
\end{equation}
and $\sigma_{RN}^{\rm eff}$ is the charmonium-nucleon effective cross section, which will
be explained below.
The total and elastic antiproton-neutron cross sections in Eq.(\ref{sigma_pbarN^inel}) are set equal 
to the antiproton-proton ones. The $\bar p p$ cross sections are taken from the PDG parametrization 
\cite{PhysRevD.50.1173}:
\begin{eqnarray}
   \sigma_{\bar p p}^{\rm tot}(p_{\rm lab}) &=& 
   38.4+77.6p_{\rm lab}^{-0.64}+0.26\ln^2(p_{\rm lab})-1.2\ln(p_{\rm lab})~,      \label{sig_pbarp^tot} \\
   \sigma_{\bar p p}^{\rm el}(p_{\rm lab}) &=&
   10.2+52.7p_{\rm lab}^{-1.16}+0.125\ln^2(p_{\rm lab})-1.28\ln(p_{\rm lab})~,     \label{sig_pbarp^el}
\end{eqnarray}
where the beam momentum, $p_{\rm lab}$, is in GeV/c and the cross sections are in mb.

Let us next turn to the time dependence of the charmonium formation. This is
expressed via the charmonium-nucleon effective cross section,
$\sigma_{RN}^{\rm eff}(p_R,z)$,
which is a function of the charmonium momentum $p_R$ in the target nucleus rest-frame
and of the distance $z$ from the $c\bar c$-pair production point. Following Refs.
\cite{Farrar:1988me,Farrar:1989vr} we express $\sigma_{RN}^{\rm eff}$ in terms of  
a formation length $l_R$:
\begin{equation}
   \sigma_{RN}^{\rm eff}(p_R,z)
  =\sigma_{RN}(p_R) \left(\left[ \left(\frac{z}{l_R}\right)^\tau
    + \frac{<n^2k_t^2>}{m_R^2} \left(1-\left(\frac{z}{l_R}\right)^\tau\right) \right]
    \Theta(l_R-z) +\Theta(z-l_R)\right)         \label{sigma_RN_eff}
\end{equation}
with $\tau=1$. In Eq.(\ref{sigma_RN_eff}), $n$ is the number of hard gluons 
in the intermediate state and $<k_t^2>^{1/2} \simeq 0.35$ GeV/c is the average
transverse momentum of a quark in a hadron. Assuming that the reaction 
$\bar p p \to R$ is dominated by $ q q q + \bar q \bar q \bar q $ 
annihilation into three hard gluons \cite{Brodsky:1988xz}, we will use the value $n=3$. 
The formation lengths of hadrons are model-dependent. 
For the $J/\Psi$ we apply a standard formula with an energy denominator 
(c.f. \cite{Farrar:1988me,Farrar:1989vr,Gerland:1998bz})
\begin{equation}
   l_{J/\Psi}     \simeq  \frac{2p_{J/\Psi}}{m_{\Psi^\prime}^2-m_{J/\Psi}^2}~. \label{l_JPsi}
\end{equation}
For $\Psi^\prime$ we rely on the estimate of Ref. \cite{Gerland:1998bz}
\begin{equation}
   l_{\Psi^\prime} \simeq  6\mbox{fm}\frac{p_{\Psi^\prime}}{30\mbox{GeV}}~,      \label{l_PsiPrime}
\end{equation}
which, however, has large theoretical uncertainty.

Formula (\ref{Gamma_pbar2R}) for the partial width of the antiproton  
with respect to the process $\bar p p \to R$ can be simplified in the 
limit of small width of the resonance $R$. To this aim 
we perform the integration over proton momentum in (\ref{Gamma_pbar2R}) 
using the spherical coordinate system with $z$-axis along the antiproton 
beam momentum:
\begin{equation}
   \Gamma_{\bar p \to R}=\frac{3\Gamma_{R \to {\bar p} p}}{2}
   \int\limits_0^{p_{F,p}}\,dp p^2\int\limits_{-1}^1\,d\cos\Theta\,
   \frac{v_{\bar p p}}{q^2} \sqrt{s} {\cal A}_R(s)~,        \label{Gamma_pbar2R_theta}
\end{equation}
where $\cos\Theta=p_z/p$, and 
$
   s=(E_p+E_{\bar p})^2 - p^2 - p_{\rm lab}^2 - 2 p p_{\rm lab}\cos\Theta
$.
It is convenient to make the angular dependence in the spectral function explicit:
\begin{equation}
   {\cal A}_R(s)= \frac{\gamma}{4 \pi p p_{\rm lab} [(A(p)-\cos\Theta)^2+\gamma^2/4]}~,
                                   \label{A_R_theta} 
\end{equation}
where $A(p)=[(E_p+E_{\bar p})^2 - p^2 - p_{\rm lab}^2 - m_R^2] / 2 p p_{\rm lab}$
and $\gamma=\sqrt{s} \Gamma_R / p p_{\rm lab}$.
If $\gamma \ll 1$, one can replace the Breit-Wigner distribution (\ref{A_R_theta})
by the $\delta$-functional distribution
\footnote{
Since $p \simeq p_{F,p} \simeq 0.3$ GeV/c, we obtain the estimate
$\gamma \sim 10^{-4}$ for the $J/\Psi$ and $\Psi^\prime$ charmonium states.
We have also checked in some selected cases that the direct (however, extremely CPU-time consuming)
Monte-Carlo calculation of the momentum integral in Eq.(\ref{Gamma_pbar2R})
gives results indistinguishable from those obtained assuming zero width of charmonium states.}
\begin{equation}
   {\cal A}_R(s) \simeq \frac{1}{2 p p_{\rm lab}}
   \delta(A(p)-\cos\Theta)~,                           \label{A_R_delta} 
\end{equation}
and set $\sqrt{s} \simeq m_R$ in Eq. (\ref{Gamma_pbar2R_theta}).
Substituting Eq.(\ref{A_R_delta}) in Eq.(\ref{Gamma_pbar2R_theta}) and performing
the integration over $\cos\Theta$ we obtain the following formula:
\begin{equation}
   \Gamma_{\bar p \to R} =
   \frac{3 m_R \Gamma_{R \to {\bar p} p}}{4 p_{\rm lab} q_R^2}
   \int\limits_{\min(p_1,p_{F,p})}^{\min(p_2,p_{F,p})}\,dp\, 
     p\, v_{\bar p p}~,                  \label{Gamma_pbar2R_theta_delta}
\end{equation}
where $q_R=\sqrt{m_R^2/4-m^2}$. The limiting momenta, $p_1$ and $p_2$,
are, respectively, the smaller and larger solutions of the equation 
$A(p)=\pm 1$.

In order to proceed further, we have to specify the dispersion relation
between the energy and momentum of a proton in the target nucleus.
The choice consistent with a model where nucleons carry all nucleus momentum in 
the infinite momentum frame is to set the proton energy constant independent 
on the proton momentum, i.e. $E_p=m-B$, where $B \simeq 8$ MeV 
is the nucleus binding energy per nucleon 
(in actual calculations we used the nucleus-dependent 
empirical values of the binding energies from Ref. \cite{Audi:2002rp}).
In this case, using expression $v_{\bar p p}=q_R m_R/E_{\bar p}E_p$ allows us
to take the momentum integral in Eq.(\ref{Gamma_pbar2R_theta_delta}) analytically:
\begin{equation} 
  \Gamma_{\bar p \to R} = 
   \frac{3 m_R^2 \Gamma_{R \to {\bar p} p}}{8 p_{\rm lab} E_{\bar p} E_p q_R}
   \left[\min(p_2,p_{F,p})^2-\min(p_1,p_{F,p})^2\right]~,      \label{Gamma_realistic}
\end{equation}
with $p_{1,2} = |p_{\rm lab} \mp \sqrt{(E_{\bar p}+ E_p)^2-m_R^2}|$.
Since $p_2 \gg p_{F,p}$, we can replace the upper integration limit
in Eqs.(\ref{Gamma_pbar2R_theta_delta}),(\ref{Gamma_realistic}) by $p_{F,p}$.
At the beam momentum of the on-shell $R$ production on the proton in vacuum at rest,
$p_1\simeq0$ and Eq.(\ref{Gamma_realistic}) simplifies to
\begin{equation}
   \Gamma_{\bar p \to R}^{\rm on-shell} \simeq
   \frac{3 m_R^2 \Gamma_{R \to {\bar p} p} p_{F,p}^2}{8 p_{\rm lab} E_{\bar p} E_p  q_R}~.
                                          \label{Gamma_onshell}
\end{equation}
Thus $\Gamma_{\bar p \to R}^{\rm on-shell} \propto \rho_p^{2/3}$. The deviation from
the usual linear density dependence originates from the narrowness of the
resonance state $R$: Owing to the Fermi motion it is difficult to find a proton which
exactly matches the on-shell resonance kinematics.  

Equation (\ref{Gamma_onshell}) leads to the estimate of Ref. \cite{Farrar:1989vr} for the 
ratio
\begin{equation}
 \frac{\sigma_{\bar p A \to R (A-1)^*}}{Z\sigma_{\bar p p \to R}}
 \simeq \frac{3 \pi m_R m \Gamma_R}{4(m_R^2-2m^2)v_{\bar p}p_{F,p}} \sim 10^{-4}~. \label{estim}
\end{equation}
for $\Gamma_R \simeq 93$ keV in the case of $J/\Psi$. Such a strong reduction implies, however,
the antiproton energy being precisely on the $R$ on-shell peak, 
i.e. $E_{\bar p}=m_R^2/2m-m$ (or $p_{\rm lab}=m_R q_R/m$).
If the beam energy resolution $\Delta E$ does
not allow to resolve the on-shell $R$ production in the $\bar p p \to R$ reaction, i.e.
$\Delta E \gg \Gamma_R$, the r.h.s. of Eq.(\ref{estim}) should be multiplied 
by $2m\Delta E/\pi m_R \Gamma_R$ (see also Ref. \cite{Farrar:1989vr}). 
We emphasize that the beam energy resolution is strongly influencing the elementary cross section
$\sigma_{{\bar p} p \to R}$, but not the cross section on the nucleus $\sigma_{\bar p A \to R (A-1)^*}$,
since the latter changes on the rather large scale $\Delta E \sim p_{F,p} \sim 0.3$ GeV only.

\section{Results}
\label{results}

We will consider the following target nuclei: $^9$Be, $^{12}$C, $^{16}$O, $^{27}$Al,
$^{40}$Ca, $^{56}$Fe, $^{63}$Cu, $^{75}$As, $^{112,116,120,124}$Sn, $^{142}$Ce, 
$^{181}$Ta, $^{197}$Au and $^{208}$Pb. This choice is mostly motivated by the availability
of neutron density parameters \cite{Koptev:1980qj,Nieves:1993ev,Schmidt:2002py}.
  
For light nuclei $(A \leq 20)$ we use the proton and neutron density profiles 
of the harmonic oscillator model
\begin{equation}
   \rho_q(r)=\rho_q^0\left[1+a_q\left(\frac{r}{R_q}\right)^2\right]
             \exp\{-(r/R_q)^2\}~,~~~q=p,n~.     \label{rhoHO}
\end{equation}
For heavy nuclei $(A > 20)$ we apply the two-parameter Fermi distributions
\begin{equation}
   \rho_q(r)=\rho_q^0
             \left[\exp\left(\frac{r-R_q}{a_q}\right)+1\right]^{-1}~,~~~q=p,n~.   \label{rho2pF}
\end{equation}
The normalization constants $\rho_q^0$ are chosen such that
\begin{equation}
   \int\,d^3 r \rho_p(r)=Z~,~~~~\int\,d^3 r \rho_n(r)=A-Z~.   \label{normCond}
\end{equation}
The charge-density-distribution parameters are taken from a standard compilation \cite{DeJager:1974dg}.
The neutron density parameters for most of nuclei are taken from Nieves et al. 
\cite{Nieves:1993ev}, who report the fits to the Hartree-Fock calculations with the density-matrix 
expansion \cite{Negele:1972zp}. In the present calculations we use the {\it point} density parameters
of protons and neutrons which were obtained from charge density parameters and 
neutron matter density parameters by employing the correction formulas 
from Ref. \cite{Nieves:1993ev}.   

The neutron density parameters for some of nuclei used in our calculations are, however, not given
in Ref. \cite{Nieves:1993ev}. For $^9$Be and $^{181}$Ta we rely upon the Glauber model analysis of
1 GeV proton charge-exchange scattering by Koptev et al \cite{Koptev:1980qj}, while for Sn isotopes
we employ the results of the antiprotonic X-ray analysis of Schmidt et al \cite{Schmidt:2002py}.
The neutron density parameters of these nuclei are collected in Table~\ref{tab:rho_n_par}.

\begin{table}[htb]
\caption{\label{tab:rho_n_par} Neutron density parameters (in fm) for some of the nuclei
used in calculations. For $^9$Be and $^{181}$Ta nuclei, the neutron matter density
while for Sn isotopes the point neutron density parameters are given.}
\begin{ruledtabular}
\begin{tabular}{ccc}
Nucleus & $R_n$ & $a_n$ \\
\hline
$^9$Be     & 2.11  & 1.000 \\
$^{181}$Ta & 6.42  & 0.640 \\
$^{112}$Sn & 5.416 & 0.543 \\
$^{116}$Sn & 5.399 & 0.552 \\
$^{120}$Sn & 5.356 & 0.565 \\
$^{124}$Sn & 5.530 & 0.558 \\
\end{tabular}
\end{ruledtabular}
\end{table} 

\begin{figure}
\includegraphics[scale = 0.6]{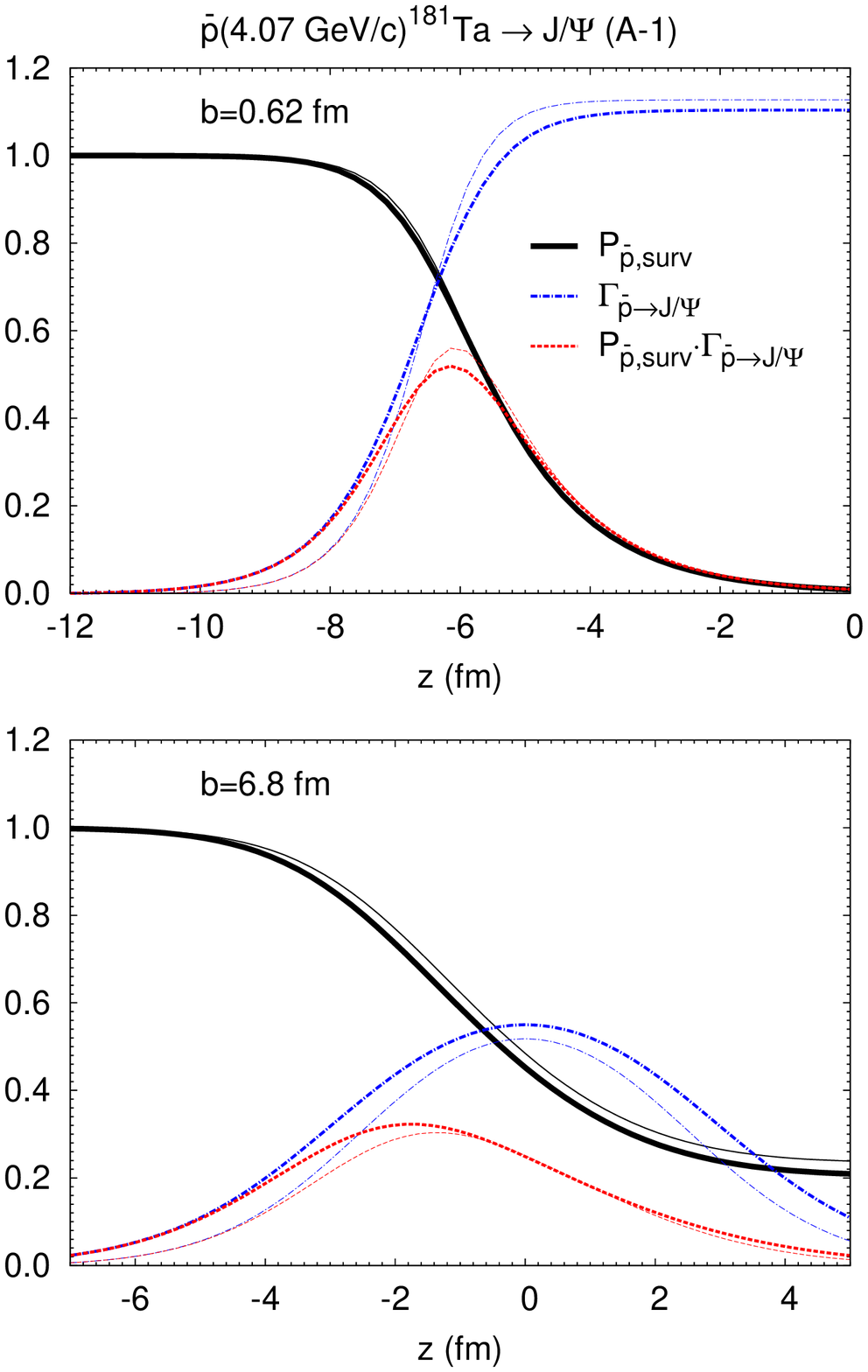}
\caption{\label{fig:zdep_Ta181}(Color online) Dependence of the antiproton
survival probability (Eq.(\ref{P_pbar_surv}), solid lines),
of the $J/\Psi$ production width
(Eq.(\ref{Gamma_onshell}), dash-dotted lines)
and of their product (dashed lines) on the longitudinal coordinate $z$
for the central ($b=0.62$ fm, upper panel) and peripheral 
($b=6.8$ fm, lower panel) collisions $\bar p+^{181}$Ta at 
$p_{\rm lab}=4.07$ GeV/c. 
The width is given in units of $10^{-8}$c/fm.
The thick lines are obtained with the charge density diffuseness parameter
$a_{\rm ch}=0.64$ fm, while the thin lines -- with $a_{\rm ch}=0.52$ fm.
The center of the nucleus is at $b=0$, $z=0$. The antiproton propagates
in the positive $z$ direction.}
\end{figure}
The main features of the antiproton-nucleus interaction with a heavy nucleus
leading to the exclusive $J/\Psi$ production are visualized in 
Fig.~\ref{fig:zdep_Ta181}, which shows the $\bar p$ survival probability 
$P_{\bar p,{\rm surv}}$, the partial $J/\Psi$ production width 
$\Gamma_{\bar p \to J/\Psi}$ and their product as functions of $z$ 
at the two different values of an impact parameter for the
$^{181}$Ta target. The beam momentum $4.07$ GeV/c is chosen to
set the produced $J/\Psi$ on-shell for the proton target at rest.
Thus, according to Eq.(\ref{Gamma_onshell}), 
$\Gamma_{\bar p \to J/\Psi} \propto \rho_p^{2/3}$.
As expected, the antiproton is almost completely absorbed in the
diffuse surface region, where the partial width $\Gamma_{\bar p \to J/\Psi}$
is relatively small. Therefore, $\bar p$-absorption strongly ($\sim 5$ times)
reduces the $J/\Psi$ production governed by the product 
$P_{\bar p,{\rm surv}}\Gamma_{\bar p \to J/\Psi}$.
Moreover, the surface absorption of $\bar p$ leads to the significant sensitivity
of $J/\Psi$ production to the diffuseness of the proton density distribution,
in-particular, for peripheral collisions (cf. thick and thin lines in
the lower panel of Fig.~\ref{fig:zdep_Ta181}).
\begin{figure}
\includegraphics[scale = 0.6]{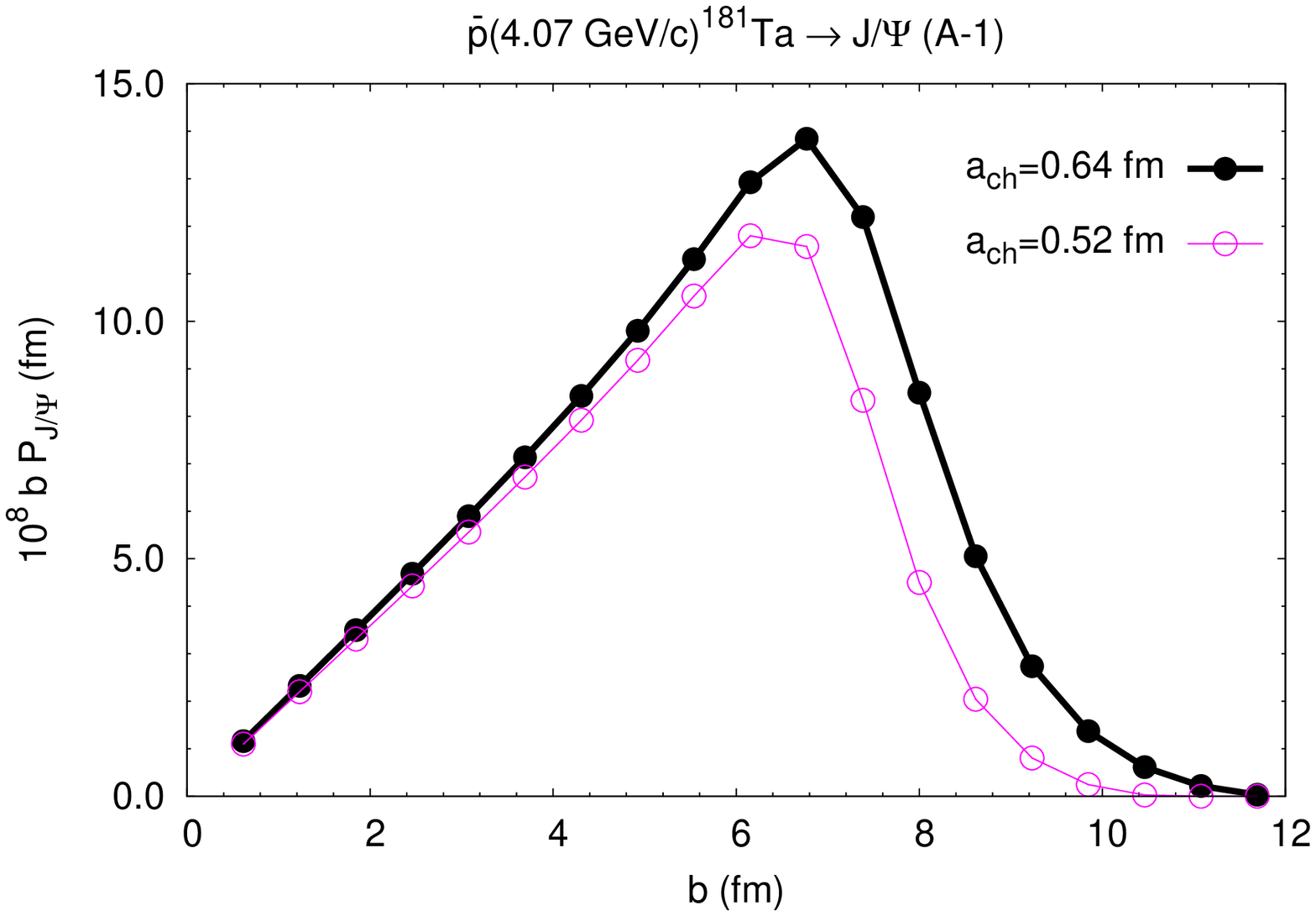}
\caption{\label{fig:Nj_Ta181}(Color online) $J/\Psi$ production
probability (Eq.(\ref{P_JPsi})) multiplied by impact parameter
$b$ as a function of $b$ for
$\bar p+^{181}$Ta collisions at $p_{\rm lab}=4.07$ GeV/c.
The lines with solid (open) circles are calculated with
the charge density diffuseness parameter $a_{\rm ch}=0.64$ 
(0.52) fm. $J/\Psi$ absorption is turned off.}
\end{figure}
This sensitivity is more clearly demonstrated in Fig.~\ref{fig:Nj_Ta181}, 
where the impact parameter dependence of the $J/\Psi$ production probability
(neglecting $J/\Psi$ absorption)
\begin{equation}
   P_{{\rm J}/\Psi}(b)=v_{\bar p}^{-1}\int\limits_{-\infty}^\infty\,dz 
   P_{\bar p,{\rm surv}}(z,b) \Gamma_{\bar p \to {\rm J}/\Psi}(z,b) \label{P_JPsi}
\end{equation}
is shown for the two slightly different values of the charge density diffuseness
parameter. The same effect shows up also in the mass dependence of
the $J/\Psi$ transparency ratio (upper panel of Fig.~\ref{fig:pbarA_syst} below).
Let us  now discuss the impact-parameter integrated cross sections. 

All cross sections have been calculated assuming the minimum bias triggering 
condition for the $\bar p$-nucleus collisions.
Numerically, this has being done by setting the upper limit for the impact
parameter integration in Eq. (\ref{sigma_pbarA2R}) equal to a large value,
$R_n+10a_n$ for all nuclei except $^9$Be, and 8 fm for $^9$Be.
    
\begin{figure}
\includegraphics[scale = 0.6]{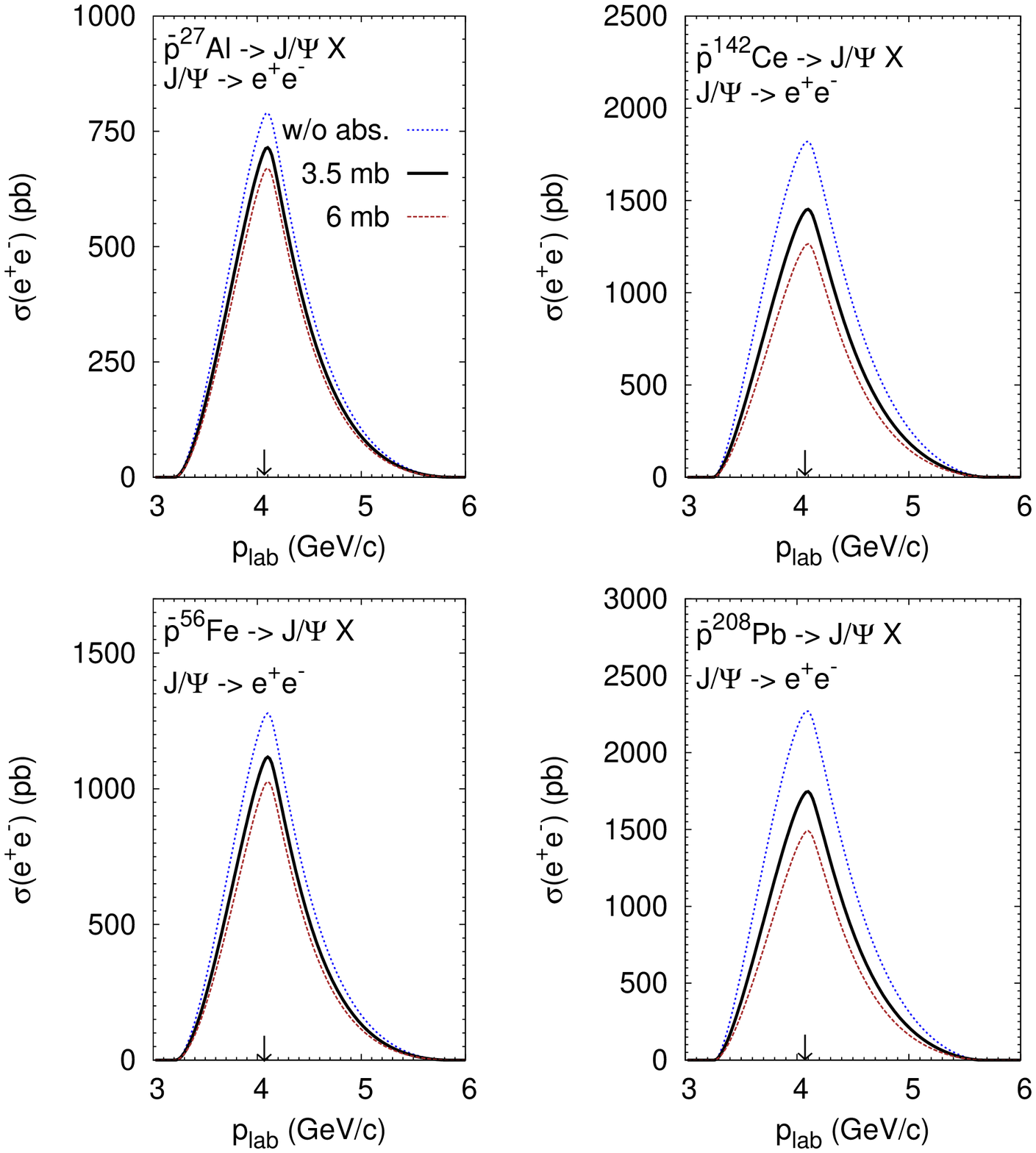}
\caption{\label{fig:pbarAlPb}(Color online) The $J/\Psi$ production cross section
in $\bar p$ collisions with $^{27}$Al, $^{56}$Fe, $^{142}$Ce and $^{208}$Pb
vs antiproton beam momentum calculated with $\sigma_{J/\Psi N}=0$ (dotted line)
$\sigma_{J/\Psi N}=3.5$ mb (solid line) and $\sigma_{J/\Psi N}=6$ mb
(dashed line). Vertical arrows show the beam momentum 4.07 GeV/c of
the on-shell $J/\Psi$ production in vacuum.}
\end{figure}
Fig.~\ref{fig:pbarAlPb} shows the beam momentum dependence of $J/\Psi$ production 
cross section for several target nuclei. Apart from the result without $J/\Psi$ 
absorption, we present two calculations with different $J/\Psi N$ dissociation 
cross sections. The choice $\sigma_{J/\Psi N} \simeq 3.5$ mb is motivated by the
early experiment on $J/\Psi$ photoproduction at $E_\gamma=20$ GeV at SLAC
\cite{Anderson:1976hi}, while the value $\sigma_{J/\Psi N} \simeq 6$ mb is obtained
in \cite{Gerschel:1993uh} from the global Glauber fit of the $J/\Psi$ 
transparency ratios in high-energy $\gamma$-, $p$-, $\bar p$- and $\pi$-induced 
reactions. The large $J/\Psi N$ inelastic cross section in the range $6-8$ mb is
reported in recent calculations employing effective Lagrangians
of the local hidden gauge theory \cite{Molina:2012mv}.
Since the $J/\Psi$-formation length $l_{J/\Psi} \simeq 0.4$ fm
at $p_{\rm lab}=4$ GeV/c, the results are practically insensitive to the
formation length effects, and we show only calculations with 
$l_{J/\Psi}=0$ in Eq.(\ref{sigma_RN_eff}).
On the other hand, the final $J/\Psi$ yield reveals a
clear sensitivity to the $J/\Psi N$ dissociation cross section which
becomes more pronounced for heavier targets.

\begin{figure}
\includegraphics[scale = 0.6]{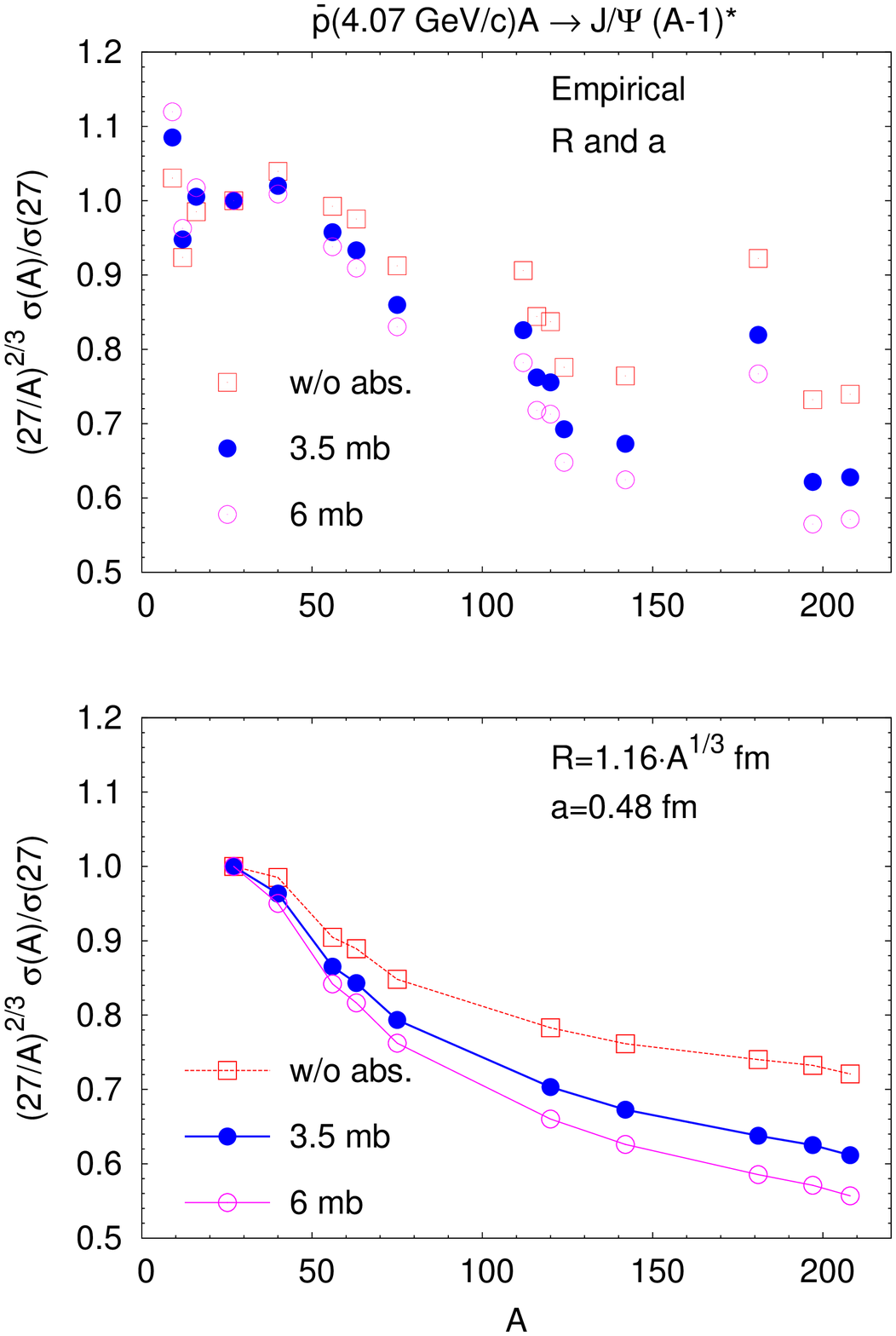}
\caption{\label{fig:pbarA_syst}(Color online) The transparency ratio of
$J/\Psi$ production in antiproton-induced reactions (\ref{S^tilde_A}) for 
the nuclei $^9$Be, $^{12}$C, $^{16}$O, $^{27}$Al,
$^{40}$Ca, $^{56}$Fe, $^{63}$Cu, $^{75}$As, $^{112,116,120,124}$Sn, $^{142}$Ce, 
$^{181}$Ta, $^{197}$Au and $^{208}$Pb plotted versus the mass number of 
the target nucleus. The ratio is normalized on 1 for $^{27}$Al.
Open squares, solid and open circles represent
calculations with $\sigma_{J/\Psi N}=0$, 3.5 mb, and 6 mb, respectively.
Upper panel -- results with density parameters $R_q, a_q$ ($q=n,p$) determined 
for each nucleus separately as described at the beginning of sec. \ref{results}.
Lower panel -- results for heavy nuclei excluding $^{112,116,124}$Sn 
with $R_n=R_p=1.16 A^{1/3}$ fm and $a_n=a_p=0.48$ fm.}
\end{figure}
In Fig.~\ref{fig:pbarA_syst} we show the transparency ratio,
\begin{equation}
   \tilde{S}_A=\frac{\sigma_{\bar p A \to R (A-1)^*}}{\sigma_{\bar p\,^{27}{\rm Al} \to R\,^{26}{\rm Mg}^*}}
       \left(\frac{27}{A}\right)^{2/3}~,    \label{S^tilde_A}
\end{equation}
calculated with the $J/\Psi$ production cross sections at their peak values 
(see Fig.~\ref{fig:pbarAlPb}) and rescaled by $A^{-2/3}$. This rescaling factor
corresponds to the surface-dominated $\bar p$ absorption at moderate
beam momenta. The nucleus $^{27}$Al is chosen for normalization, since this is 
the lightest one in our set of 
selected nuclei which has the two-parameter Fermi density 
distributions of nucleons. Being defined in this way, the transparency ratio better 
represents the systematic mass dependence for heavy nuclei.

The transparency ratio $R_A$ reveals strong local variations as a function of 
the mass number when calculated with the empirical nucleon density parameters
(upper panel of Fig.~\ref{fig:pbarA_syst}).
This arises from the details of empirical density profiles.
For example, the local maximum for the $^{181}$Ta nucleus appears due to the large 
diffuseness parameter of the charge
distribution, $a_{\rm ch}=0.64$ fm \cite{DeJager:1974dg}.
These local variations, as expected, disappear if we enforce the density profiles 
to be determined by the uniform parameters (lower panel of Fig.~\ref{fig:pbarA_syst}).

Another peculiar feature observed in Fig.~\ref{fig:pbarA_syst} (upper panel)
is a strong drop
of the transparency ratio along the isotope chain $^{112-124}$Sn.
If we turn off absorption of both $\bar p$ and $J/\Psi$, then
the cross section for $J/\Psi$ production at the on-shell peak
varies along this isotope chain by about $5\%$ only, because the 
proton density distribution is similar for the different 
isotopes. Therefore, this drop is mostly caused by 
$\bar p$ absorption and $J/\Psi$ dissociation on the neutron excess 
in heavier isotopes.

The sensitivity of the transparency ratio $R_A$ to the input
$J/\Psi N$ dissociation cross section is clearly visible in
Fig.~\ref{fig:pbarA_syst}.
\begin{figure}
\includegraphics[scale = 0.6]{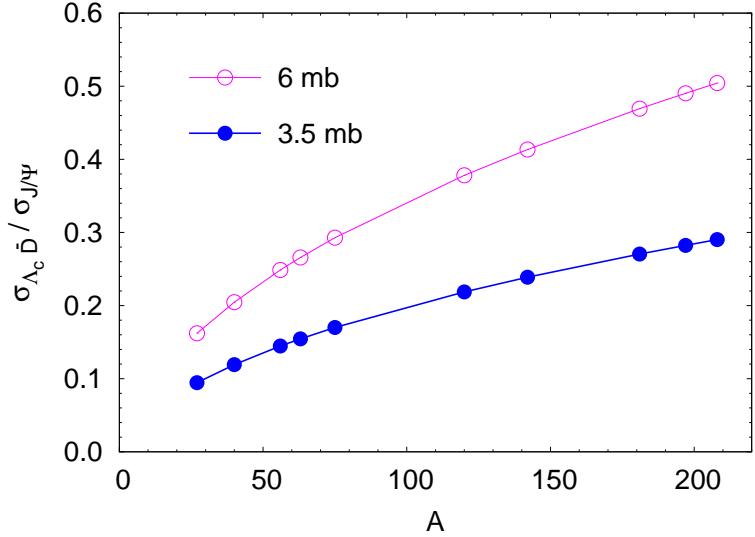}
\caption{\label{fig:Lambdac_vs_A}(Color online) 
The ratio of
the $\Lambda_c \bar D$ production cross section to the $J/\Psi$
production cross section at $p_{\rm lab}=4.07$ GeV/c
for $\sigma_{J/\Psi N}=3.5$ mb (full blue circles) 
and for $\sigma_{J/\Psi N}=6$ mb (open purple circles).
Heavy nuclei excluding $^{112,116,124}$Sn are shown 
using $R_n=R_p=1.16 A^{1/3}$ fm and $a_n=a_p=0.48$ fm.}
\end{figure}
The only possible strong interaction 
channels of the $J/\Psi$ dissociation on a nucleon below 
$D \bar D$ production threshold ($p_{\rm thr}=5.18$ GeV/c) are
$J/\Psi N \to \Lambda_c \bar D$ + up to three pions. 
Hence, the $J/\Psi$ dissociation cross section at the
beam momentum of $\simeq 4$ GeV/c is equal to the inclusive
$\Lambda_c \bar D$ production cross section on the nucleon.
Moreover (see also \cite{Gerland:2005ca}), both $\Lambda_c$
and $\bar D$ can not be absorbed in a nucleus for 
$p_{\rm lab} \simeq 4$ GeV/c  but can only change momenta
by rescattering on nucleons. For $\bar D$, the absorption
is actually forbidden for any momentum by charm conservation in strong 
interactions. For $\Lambda_c$, the threshold momentum
in the nucleon rest frame for the $\Lambda_c N \to  N N D$
is 3.55 GeV/c, while the maximum $\Lambda_c$ momentum 
in $J/\Psi N \to \Lambda_c \bar D$ is 3.34 GeV/c.
The direct channels $\bar p p \to D \bar D$ ($p_{\rm thr}=6.45$ GeV/c)
and $\bar p p \to \Lambda_c \bar \Lambda_c$ ($p_{\rm thr}=10.16$ GeV/c)
are not reachable at the $J/\Psi$ production threshold. 
Therefore, the cross section of the $\Lambda_c \bar D$-pair production
in $\bar p A$ collisions at $p_{\rm lab} \simeq 4$ GeV/c can be simply
calculated as 
\begin{equation}
   \sigma_{\Lambda_c \bar D}=\sigma_{\bar p A \to J/\Psi (A-1)^*}^{w/o\,J/\Psi abs.}
    - \sigma_{\bar p A \to J/\Psi (A-1)^*}~,   \label{sigma_LcDbar}
\end{equation}
where $\sigma_{\bar p A \to J/\Psi (A-1)^*}$ is given by Eq. (\ref{sigma_pbarA2R})
and $\sigma_{\bar p A \to J/\Psi (A-1)^*}^{w/o\,J/\Psi abs.}$ 
-- by the same Eq. (\ref{sigma_pbarA2R}), but with $P_{J/\Psi,{\rm surv}}=1$.
In Fig.~\ref{fig:Lambdac_vs_A} we show the ratio 
$\sigma_{\Lambda_c \bar D}/\sigma_{\bar p A \to J/\Psi (A-1)^*}$ at the
on-shell peak of the $J/\Psi$ production vs target mass number.
One sees the strong sensitivity of this ratio to the assumed value
of the $J/\Psi N$ dissociation cross section.
  
\begin{figure}
\includegraphics[scale = 0.6]{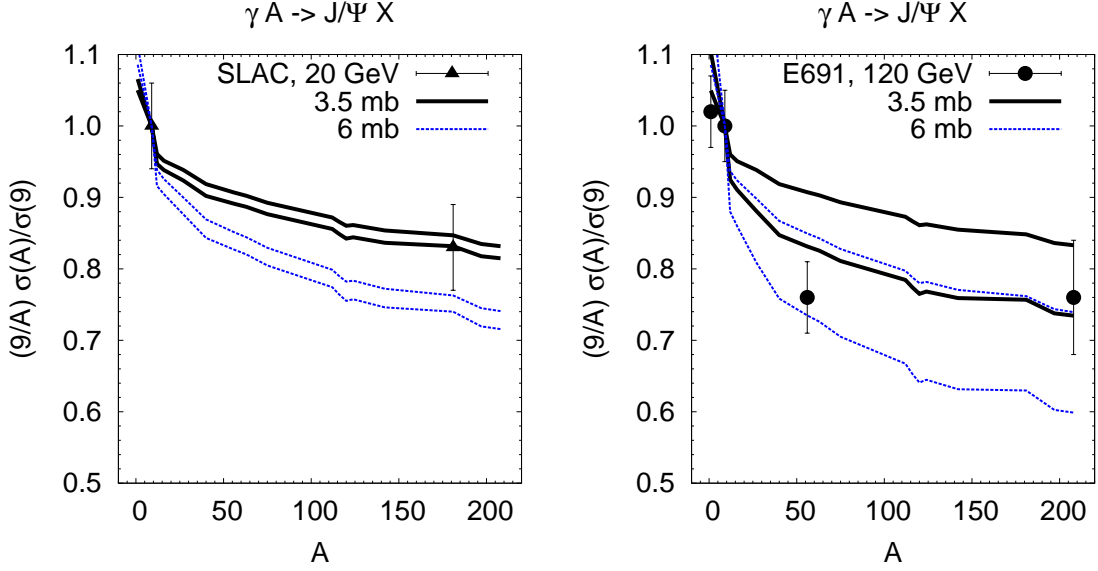}
\caption{\label{fig:gammaA}(Color online) Transparency ratio $S_A$
for the $J/\Psi$ production in $\gamma$-induced reactions on nuclei
vs target mass number. Left panel -- $E_\gamma=20$ GeV.
Right panel -- $E_\gamma=120$ GeV.
The results with $\sigma_{J/\Psi N}=3.5$ mb and  $\sigma_{J/\Psi N}=6$ mb
are shown by the solid and dashed lines, respectively.
The lower and upper lines correspond to the calculations with 
formation length $l_{J/\Psi}=0$ and $l_{J/\Psi}=2~(12)$ fm
for $E_\gamma=20~(120)$ GeV. 
Experimental data points from SLAC at $E_\gamma=20$ GeV ($^9$Be and $^{181}$Ta targets) 
\cite{Anderson:1976hi}
and from Fermilab at $E_\gamma=120$ GeV (p, $^9$Be, $^{56}$Fe, and $^{208}$Pb targets)
\cite{Sokoloff:1986bu} represent the incoherent $J/\Psi$ photoproduction cross section per nucleon
normalized on 1 for $^9$Be.}
\end{figure}
We recall that the formation time effects
are almost negligible and do not create an additional
ambiguity for the $J/\Psi$ production in low-energetic
antiproton-nucleus reactions.
In contrast, formation time effects are very important
for the $\gamma$-induced $J/\Psi$ production on nuclei.
The transparency ratio in $\gamma$-induced reactions 
is defined according to Eq.(\ref{S_A}), which in the simplest approximation
is expressed as (c.f. \cite{Gerschel:1993uh}):
\begin{equation}
   S_A = \frac{\sigma_{\gamma A \to J/\Psi X}}{A\sigma_{\gamma p \to J/\Psi X}}
       = \frac{2 \pi}{A} \int\limits_0^\infty\,db\, b\,
         \int\limits_{-\infty}^\infty\,dz \rho(z,b) 
         P_{J/\Psi,{\rm surv}}(z,b)~.                 \label{S_A_gamma}
\end{equation}
This expression, however, is valid only at low photon energies, i.e.
when the coherence length $l_c=2E_\gamma/m_{J/\Psi}^2$ is much less than
the nuclear radius.
For $E_\gamma=20$ GeV and $120$ GeV, where the $J/\Psi$ production in photon-induced reactions 
is measured \cite{Anderson:1976hi,Sokoloff:1986bu}, the coherence length is already quite large, 
$l_c=0.8$ fm and $4.8$ fm, respectively. The deviations from the classical probabilistic
formula (\ref{S_A_gamma}) appear in the Glauber model due to the coherent addition of the
production amplitudes on the two nucleons separated by the distance less than $l_c$
\cite{Bauer:1977iq,Hufner:1996dr}. In Refs. \cite{Bauer:1977iq,Hufner:1996dr},
the formulas have been derived which generalize Eq.(\ref{S_A_gamma}) for arbitrary
values of $l_c$ (c.f. Eqs. (4.2a),(4.2b) in Ref.\cite{Bauer:1977iq}
and Eq. (13) in Ref. \cite{Hufner:1996dr}). The similar expressions are also given
in Refs. \cite{Hufner:1996jw,Ivanov:2002kc}. Although these expressions somewhat differ from
each other (mainly because of the different assumptions on the vector meson -- nucleon
elastic cross section), they all give the same limits of $l_c \to 0$ and $l_c \to \infty$.
In the former case one gets Eq.(\ref{S_A_gamma}), while in the latter case one has to
replace $P_{J/\Psi,{\rm surv}}(z,b) \to P_{J/\Psi,{\rm surv}}(-\infty,b)$ in (\ref{S_A_gamma}).
We will, thus, interpolate between these two limits by simply replacing
$P_{J/\Psi,{\rm surv}}(z,b) \to P_{J/\Psi,{\rm surv}}(z-l_c,b)$ in Eq.(\ref{S_A_gamma}).
This should be a quite rough approximation, but it serves at least our purposes of the 
exploratory studies of the photoproduction.   

Fig.~\ref{fig:gammaA} shows the mass dependence of the transparency ratio
$S_A$ corrected for the coherence length effects. We present results for the two
previous values of the $J/\Psi N$ dissociation cross section, $\sigma_{J/\Psi N}=3.5$ mb 
and 6 mb. However, the charmonium formation length $l_{J/\Psi}=2$ (12) fm 
at $p_{J/\Psi}=20$ (120) GeV/c is comparable with the nuclear size.
Thus, the effective cross section $\sigma_{J/\Psi N}^{\rm eff}$ of Eq. (\ref{sigma_RN_eff})
is now substantially reduced with respect to $\sigma_{J/\Psi N}$ for 
the longitudinal coordinate within the nuclear target bulk region.
The uncertainty in the determination of the charmonium formation length has
an immediate feedback on the extraction of the genuine charmonium-nucleon
dissociation cross section. As demonstrated in Fig.~\ref{fig:gammaA},
for $E_\gamma=20$ GeV, it is still possible to clearly see the difference
between transparency ratios calculated with different values of $\sigma_{J/\Psi N}$.
However, for $E_\gamma=120$ GeV, the large formation length washes-out the sensitivity
of $S_A$ to $\sigma_{J/\Psi N}$. Moreover, the experimental errors 
do not allow to set tight constraints on $\sigma_{J/\Psi N}$.

\begin{figure}
\includegraphics[scale = 0.6]{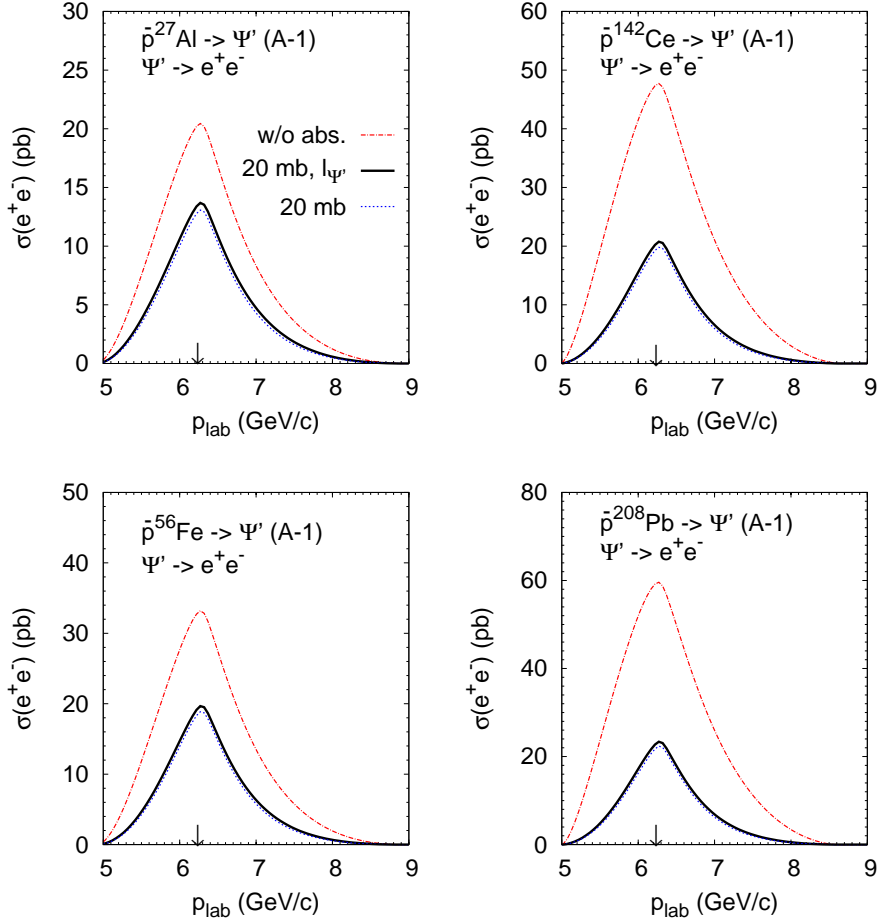}
\caption{\label{fig:pbarA_Psip_vs_plab}(Color online) $\Psi^\prime(2S)$ production
cross section for $\bar p$ collisions with $^{27}$Al, $^{56}$Fe, $^{142}$Ce and $^{208}$Pb
vs antiproton beam momentum calculated with $\sigma_{\Psi^\prime N}=0$ (dash-dotted line),
$\sigma_{\Psi^\prime N}=20$ mb \cite{Gerland:1998bz} with formation length $l_{\Psi^\prime}$
defined according to Eq.(\ref{l_PsiPrime}) (solid line) 
and with $l_{\Psi^\prime}=0$ (dotted line).
Vertical arrows show the antiproton beam momentum of the on-shell 
$\Psi^\prime(2S)$ production in vacuum, $p_{\rm lab}=6.23$ GeV.} 
\end{figure}
\begin{figure}
\includegraphics[scale = 0.6]{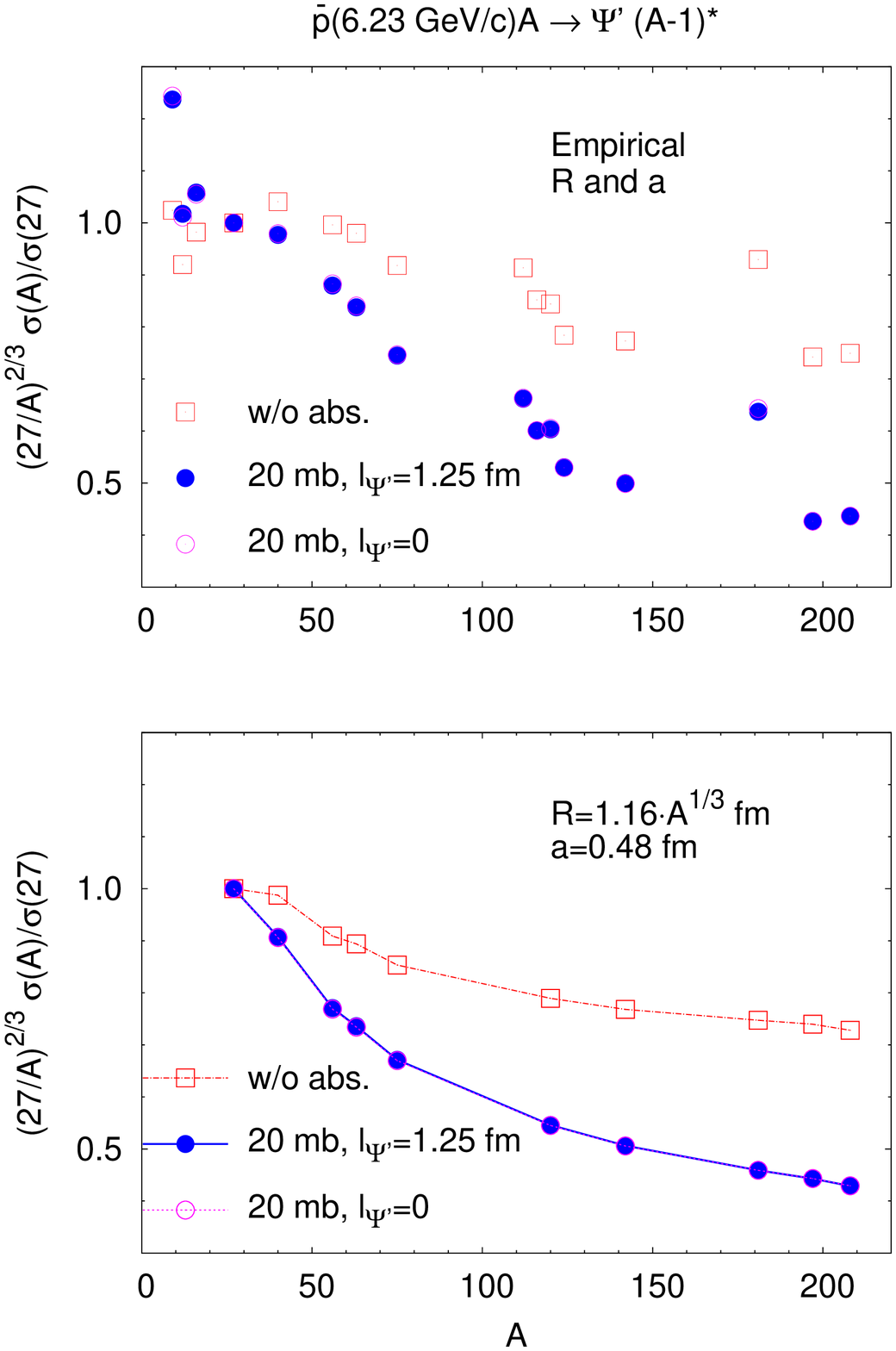}
\caption{\label{fig:pbarA_Psip_vs_A}(Color online) The transparency ratio of
$\Psi^\prime$ production in antiproton-induced reactions (\ref{S^tilde_A}) as a function
of the target mass number. Open squares show the calculation without $\Psi^\prime$   
absorption. Full and open circles represent the results with 
$\sigma_{\Psi^\prime N}=20$ mb \cite{Gerland:1998bz} with $\Psi^\prime$-formation 
length $l_{\Psi^\prime}$ given by Eq.(\ref{l_PsiPrime}) at $p_{\rm lab}=6.23$ GeV/c and
with $l_{\Psi^\prime}=0$, respectively. Upper panel -- calculations with
empirical density parameters $R_q, a_q$ ($q=n,p$) (see the beginning of sec. \ref{results}).
Lower panel -- calculations for heavy nuclei excluding $^{112,116,124}$Sn 
with uniform density parameters $R_n=R_p=1.16 A^{1/3}$ fm and 
$a_n=a_p=0.48$ fm.}
\end{figure}
Figures~\ref{fig:pbarA_Psip_vs_plab} and \ref{fig:pbarA_Psip_vs_A} present
the beam momentum and mass number dependence of $\Psi^\prime(2S)$ production 
in $\bar p$-induced reactions. Both dependences are quite similar to those
for the $J/\Psi$ production (c.f. Figs.~\ref{fig:pbarAlPb} and \ref{fig:pbarA_syst}). 
The local variations of the transparency ratio (upper panel in Fig.~\ref{fig:pbarA_Psip_vs_A})
due to the empirical density profiles are again visible. The smooth behavior of the transparency
ratio as a function of the mass number is recovered if we substitute the empirical 
density parameters by the uniform ones (lower panel in Fig.~\ref{fig:pbarA_Psip_vs_A}).  
  
To provide some hints on the possible charmonium absorption effects, 
we show in Figs.~\ref{fig:pbarA_Psip_vs_plab} and \ref{fig:pbarA_Psip_vs_A} 
the calculations with $\sigma_{\Psi^\prime N}=20$ mb as theoretically estimated 
in Ref. \cite{Gerland:1998bz}. This reduces the $\Psi^\prime$ yield 
by about a factor of 2-3 with respect to the calculation without $\Psi^\prime$ 
absorption. It is interesting to note, that with such a strong absorption, 
the $\Psi^\prime$ production becomes almost independent on the target mass number. 

Since the beam momentum is now larger than for $J/\Psi$ production, 
the formation length effects become visible (cf. solid and dotted lines 
in Fig.~\ref{fig:pbarA_Psip_vs_plab}).
However, they are still weak as compared to the uncertainty caused by the largely 
unknown $\Psi^\prime N$ cross section.

\begin{figure}
\includegraphics[scale = 0.6]{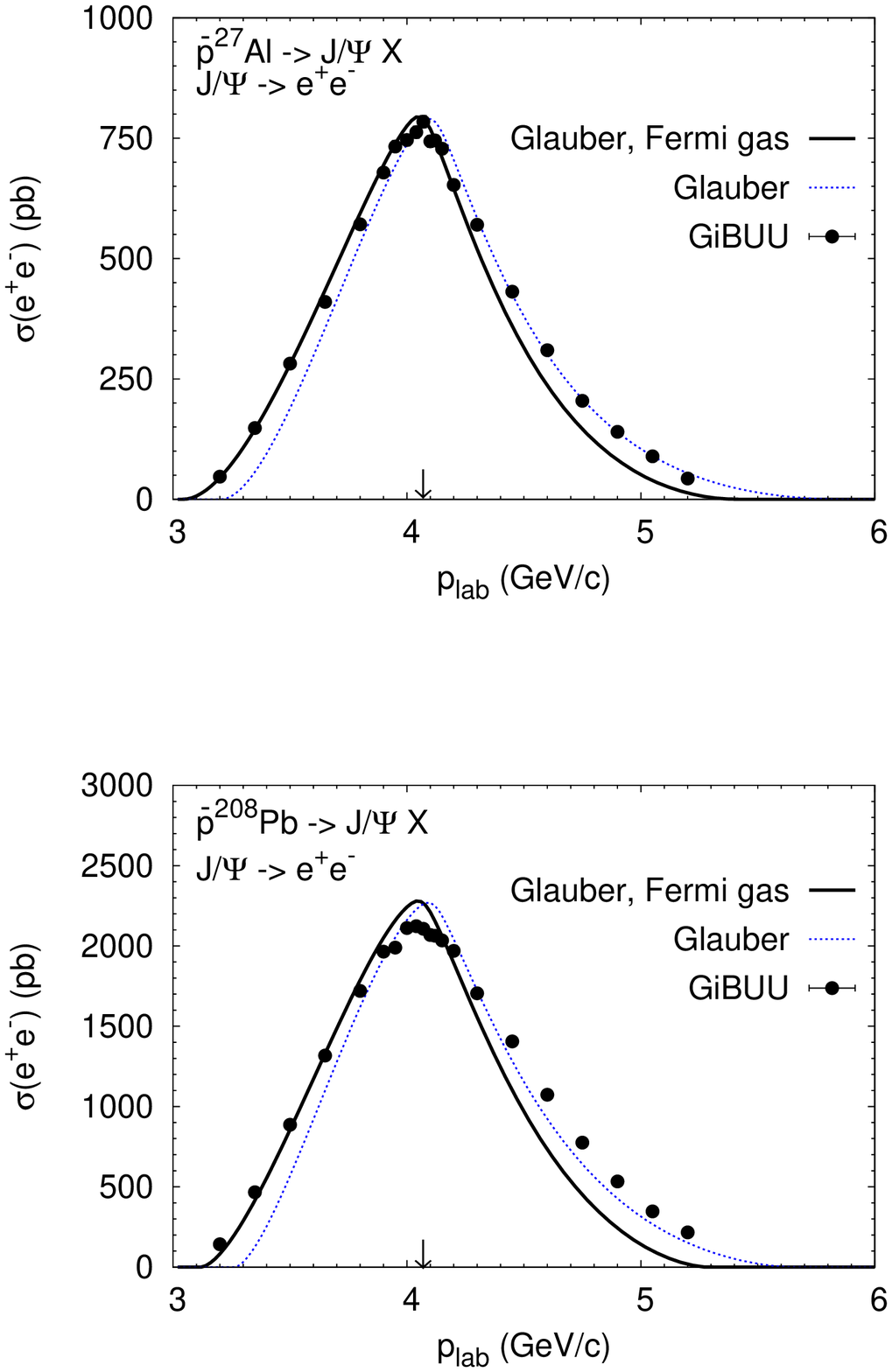}
\caption{\label{fig:GiBUU}  Comparison of the Glauber and GiBUU model calculations
for the $J/\Psi$ production cross section in $\bar p$ collisions with
$^{27}$Al and $^{208}$Pb vs beam momentum.
Dashed lines correspond to our standard Glauber model calculations with the fixed proton
energy $E_p=m-B$ as described in sect. \ref{model}. Solid lines are obtained
within the Glauber model assuming the local Fermi gas for the nuclear ground state,
same as in GiBUU. $J/\Psi$ absorption cross section is set to zero.}
\end{figure}
Finally, we address the multistep processes which are not included in the Glauber
model. For this purpose we have performed the GiBUU
model \cite{Buss:2011mx} calculations of the $\bar p+^{27}$Al and $\bar p+^{208}$Pb reactions.
The GiBUU model takes into account the annihilation as well as the elastic and inelastic 
rescattering of the incoming antiproton with empirical cross sections. 
The nuclear density profiles are chosen
to be identical in both, Glauber and GiBUU calculations. The antiproton-nucleon total
and elastic cross sections (\ref{sig_pbarp^tot}),(\ref{sig_pbarp^el}) coincide in the 
both models as well.
A comparison of the GiBUU and Glauber calculations is presented in Fig.~\ref{fig:GiBUU}. 
For simplicity, the nucleus was modeled in the local Fermi gas approximation in GiBUU.
Therefore, for the comparison purposes we have also performed the Glauber model calculations
by doing the same assumption (solid lines in Fig.~\ref{fig:GiBUU}). This has been achieved
by replacing Eq.(\ref{Gamma_realistic}) for the charmonium production width by
the following formula:
\begin{equation} 
   \Gamma_{\bar p \to R}^{FG} =
   \frac{3 m_R^2 \Gamma_{R \to {\bar p} p}}{4 p_{\rm lab} E_{\bar p} q_R}
   \left[\sqrt{\min(p_2,p_{F,p})^2+m^2}
     - \sqrt{\min(p_1,p_{F,p})^2+m^2}\right]~, \label{Gamma_FG} 
\end{equation}
where $p_{1,2}=|p_{\rm lab}(m_R^2-2m^2) \mp 2 E_{\bar p} m_R q_R|/2m^2$.
The Glauber calculation is quite close to the GiBUU results at the peak.
However, at higher beam momenta the Glauber model (with the local Fermi gas assumption)
underpredicts GiBUU somewhat.
The reason is that the fast antiproton has a chance to be decelerated
by elastic or inelastic collisions with nucleons and get momentum closer
to the peak momentum, where the cross section of $J/\Psi$ production is 
larger (see also discussion in \cite{Farrar:1989vr}).
This mechanism is taken into account in GiBUU while it is neglected 
in the Glauber model.

\section{Conclusions}
\label{conclusions}

We have performed the Glauber model calculations of $J/\Psi(1S)$ and $\Psi^\prime(2S)$ 
charmonium production in $\bar p$-nucleus collisions at $p_{\rm lab}=3-10$ GeV/c.
For the both charmonia, we have focused on the beam momentum range near the corresponding 
on-shell production peaks. Thus, only the $\bar p p \to R$ channel was taken into
account. The main nuclear effect is the broadening and reduction of the
narrow charmonium production peak due to the nuclear Fermi motion.

The $J/\Psi$ production cross section in $\bar pA$ collisions strongly depends
on the input $J/\Psi N$ dissociation cross section. This dependence is not
blurred by the charmonium formation length, in contrast
to $J/\Psi$ production in $\gamma$-induced reactions at high energies.

The surface-dominated antiproton absorption leads to relatively large local 
variations of the $J/\Psi$ transparency ratios as a function of the target mass number.
This is due to the delicate interplay between the neutron and proton density profiles:
The former governs the absorption range of the antiproton, while the latter defines 
the space region, where the $J/\Psi$ is produced. We conclude,
that the quantitative determination of the $J/\Psi N$ dissociation cross
section from experimental data on $J/\Psi$ production in $\bar pA$ reactions relies 
on the detailed and realistic description of the neutron
and proton density distributions. 
 
We would like to recall at this point that the spreading of the proton momentum distribution
due to the short-range correlations has not been taken into account in the present
calculations. Although these effects are very important at extreme kinematics regions,
they will not sensitively modify our results near the on-shell peaks of charmonium production.
The reason is that in this case the momentum integration in Eq.(\ref{Gamma_pbar2R})
is not restricted from the low-momentum side, and, hence, is only weakly sensitive
to the high-momentum tail. Another reason is that the short-range correlations
become weaker at the nuclear surface, where the antiproton is predominantly 
absorbed. Overall, the short-range correlations may create the additional 
uncertainty of $\sim 10\%$ in the cross sections of the charmonium production
close to the peak value.
The same is true for the multistep effects due to the rescattering
of the incoming antiproton in heavy target nuclei (Fig.~\ref{fig:GiBUU}).

It is expected from QCD, that the $\Psi^\prime N$ cross section is 
a factor of 2-4 larger than the $J/\Psi N$ cross section due to the larger size 
of the $\Psi^\prime$ as compared to the $J/\Psi$ and it may reach up to 20 mb  
\cite{Gerland:1998bz,Hufner:1997jg}. Such a strong absorption will be certainly
testable with the new PANDA detector at FAIR starting from 2018.
Having all the above uncertainties of our calculations in mind, we conclude that 
the measurements of the $J/\Psi$ transparency ratio with a precision of
at least $\sim 20\%$ would allow determination of the $J/\Psi$-nucleon 
dissociation cross section with accuracy of about 3 mb.

Significant $J/\Psi N$ cross section implies the corresponding enhancement 
of the $\Lambda_c \bar D$ production, since near $J/\Psi$ production threshold
$J/\Psi N \to \Lambda_c \bar D$ is the only possible
inelastic channel of $J/\Psi N$ interaction.
On the other hand, there are several models which give widely spread predictions 
for the $J/\Psi N \to \Lambda_c \bar D$ cross sections 
\cite{Sibirtsev:2000aw,Oh:2007ej,Hilbert:2007hc,Molina:2012mv}.
Thus PANDA offers an interesting possibility to test these predictions
by measuring the ratio of the $\Lambda_c \bar D$- to $J/\Psi$-production
cross sections (c.f. Fig. \ref{fig:Lambdac_vs_A}).

\begin{acknowledgments}
A.L. acknowledges useful discussions with Prof.~H.~Lenske 
and Prof.~U.~Mosel, the hospitality and financial support of the
Giessen University, where this work was started.
M.S. wants to thank Helmholtz Institute in Mainz for support during
initial stage of work on this project.
This work was supported by HIC for FAIR within the framework of
the LOEWE program (Germany), and by the Grant NSH-215.2012.2 (Russia).
\end{acknowledgments}

\bibliography{pbarCharm2}

\end{document}